# Low temperature pressureless "immediate sintering" of novel nanostructured WC/Co/NiCrSiB-alloy cemented carbide

H. Amel-Farzad[a], E. Taheri-Nassaj[a*], D. Meertens[b], R.E. Dunin-Borkowski[b], A.H. Tavabi[b]

[a]: Department of Materials Engineering, Faculty of Engineering, Tarbiat Modares University, P.O. Box 14115-143, Tehran, Iran
[b]: Ernst Ruska-Centre for Microscopy and Spectroscopy with Electrons, Forschungszentrum Jülich, 52425 Jülich, Germany

## Abstract

A novel nanostructured cemented carbide formed from WC-5%Co-20%BNi2 brazing alloy is described. During sintering, the BNi2 alloy is infiltrated into a green compact of WC-5%Co at 1050-1100 °C for 2-60 minutes. Perfect wetting behavior and a zero contact angle are achieved after only 40 s. Relative densities of 98.5% and 100% and microhardness values of above 1500HV1 and 1800HV1 are obtained after 2 and 30 minutes, respectively. A change in mean particle size of 0.6 μm in the precursor to a bimodal distribution of 350-400 nm and 10-20 nm is explained by a solution/reprecipitation mechanism.

Cemented carbides, which have been used widely as cutting tools for almost a century, are usually formed by sintering tungsten carbide (WC) particles with a metallic binder [1]. Their favorable properties, which include cold and hot hardness, toughness, bending strength and a low coefficient of thermal expansion, are typically achieved by using a Co binder to wet the WC particles [1–5]. Research into alternatives for Co is motivated by its cost, the undesirable health effect of Co particles that may be released by corrosion or wear and the low corrosion/oxidation resistance of WC-Co [6–8]. Logical alternatives include the other ferrous family elements Fe and Ni, which both exhibit solubility for WC and have good mechanical properties. The use of Fe as a binder usually leads to the formation graphite for thermodynamic reasons [9]. Ni has been shown to eliminate all of the problems mentioned above [10]. However, WC-Ni tools have lower toughness when compared to WC-Co [9], as a result of the lower stacking fault energy of Ni compared to Co [9], the lower wettability of WC by Ni than by Co [11] and/or the lower fracture strain of Ni than Co [12]. Efforts have been made to improve the wettability of WC by the Ni binder, including the addition of $Mo_2C$ to the system [12]. Another approach involves the replacement of Ni by its alloys [13]. According to phase diagrams, the addition of Si, Al, Mn, Cr, Nb and/or Fe to the Ni matrix can strengthen it by a solid solution mechanism [14]. Correa et al. [15] used a Ni-4.1Si alloy binder and reported a relative density of 98.3 - 98.8% and an enhancement in both strength and toughness. Worauaychai et al. [16] evaluated the effect of adding Cu, Sn and P to the Ni matrix on the sintering behavior of WC-Ni composites and reported a significant increase in hardness and a remarkable decrease in sintering temperature on alloying the Ni binder with less than 0.05% P (but not Cu or Sn). Both references mention the importance of fluidity for the densification and toughening of such composites. Another approach to improve both the hardness and the toughness of WC-Ni composites is refine the grain size to the nanoscale [9].

WC-based tools are typically brazed easily using Cu or Ag-based brazing alloys, not Ni-based ones. Nevertheless, Ni-based hardfacing alloys are used on the surfaces of various materials in the form of composites with WC particles [16–20]. Although some of these coatings are applied using processes that result in , it is difficult to find signs of insufficient wetting in the microstructures of such hardfaced samples [20,21]. The melting temperature and viscosity of Ni alloys are usually decreased by adding P, Si and/or B, of which P is the most effective, but forms highly brittle phosphide phases. Most grades include B and/or Si with similar effects, while Cr and Fe are sometimes also added for solid solution strengthening [22].

Here, we describe an innovative approach based on the use of classical Ni-based brazing/hardfacing alloys as binders in the WC/binder system. The high molten fluidity of such alloys makes them suitable for achieving good WC/binder wettability and, hence, higher densification and improved mechanical properties, while their low melting temperature makes the process much easier. The sintering behavior of a WC-5%wtCo cemented carbide is studied, as infiltrated by a commercial nickel-based brazing alloy (referred to as BNi2 by AWS [23]) containing B, Si, Cr and Fe, and two other similar commercial hardfacing alloys (referred to as Ni22 and Ni35A) (Table 1).

Table 1
Chemical composition and melting temperature of NiCrSiB (brazing or hardfacing) binder alloys considered in the present study.

| Sample name | Chemical composition of NiCrSiB alloys (wt%) | | | | | | | Melting range, based on DTA data (°C) | Classification |
|---|---|---|---|---|---|---|---|---|---|
| | Cr | Si | B | Fe | C | Mn | Ni | | |
| Ni22 | 1.0 | 0.0 | 1.4 | 1.0 | 0.2 | 0.1 | 96.3 | 853-874 | Hardfacing alloy |
| Ni35A | 10 | 3.0 | 2.2 | 5.0 | 0.0 | 0.0 | 79.8 | 856-865 | Hardfacing alloy |
| BNi2 | 7.0 | 4.5 | 3.0 | 3.0 | 0.0 | 0.0 | 84.5 | 957-964 | Brazing alloy |

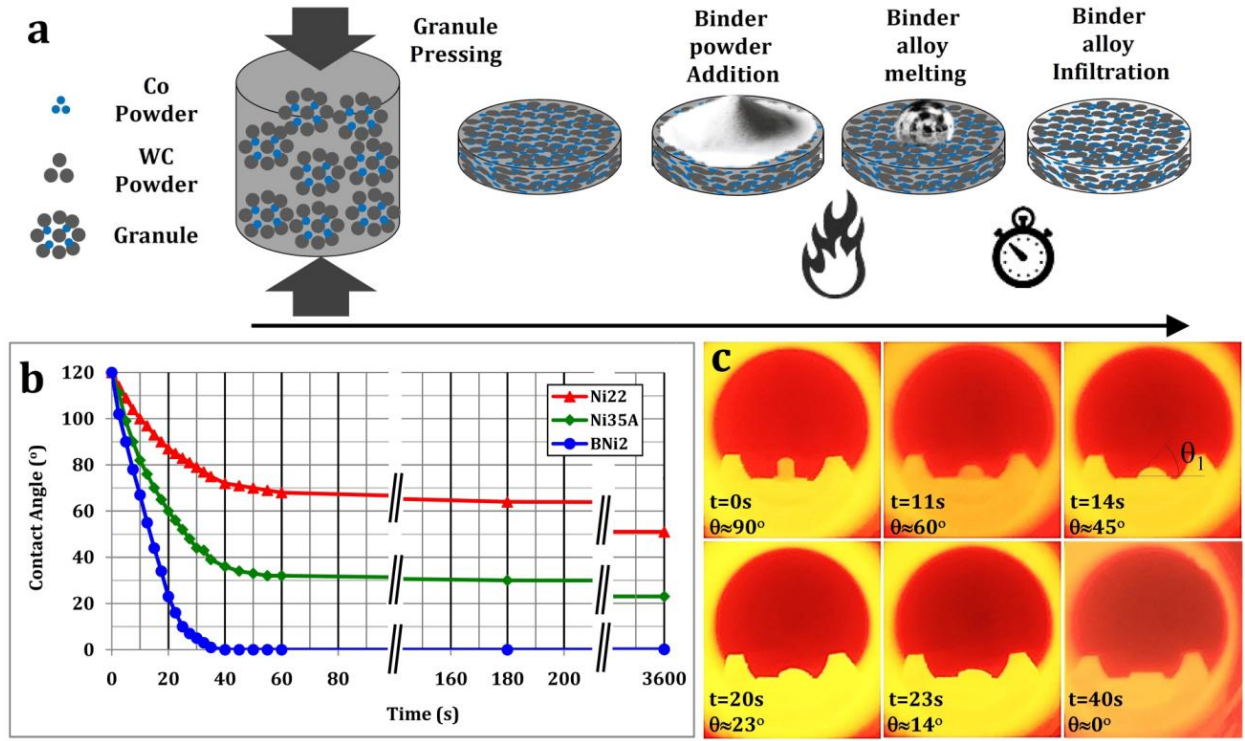

Fig. 1. (a) Schematic diagrams of the preparation and sintering process described in the text. (b) Variation in contact angle of three NiCrSiB binder alloy drops on green K05 disks plotted as a function of sintering time at 1080 °C. The powders on the sample melt to a drop with a contact angle of almost 120° after 30-35 s at 1080 °C, which is considered as the beginning of the test. (c) Consecutive frames showing the wetting behavior of a green K05 disk by a liquid BNi2 rod with time, leading to perfect wettability after 40 s.

Three commercial Ni-based (NiCrSiB) brazing or hardfacing alloy powders with various B, Si and Cr contents and different molten fluidities were used as a binder, in addition to Co (Table 1). Industrial granules (d50 = 250 μm) of submicron WC - 5%wtCo (d50$_{WC}$ = 0.6 μm), referred to as K05 by ISO, were also used. The K05 granules were pressed, to a relative density of 60%, into a disk shape with a thickness of 3 mm and a diameter of 10 mm (Fig. 1a). They were placed onto an alumina plate (with no wettability by the Ni-alloys). Weighed amounts of the Ni-alloy (8-20wt%) powders were placed on its top surface for inward infiltration into the samples (Fig. 1a). The sintering temperatures were selected on the basis of DTA results obtained from the three NiCrSiB alloys at a heating rate of 6 °C/min in 5-N purity $N_2$ gas. As listed in Table 1, all three alloys have melting temperatures of well below 1000 °C. The brazing alloy, BNi2, melts at about 957-964 °C, while the hardfacing alloys melt earlier at about 860 °C. However, they are all usually recommended for use at above 1050 °C [22]. It should be noted that parameters such as the partial pressure of $N_2$ and $O_2$ in the atmosphere, the heating rate and surface oxidation can all affect the working temperatures of these alloys [22]. It was observed that none of them could flow and infiltrate into the K05 green compacts at temperatures below 1050 °C. The sintering cycles were carried out in a low vacuum ($10^{-1}$ Torr) tube furnace after evacuating and washing the tube with 5N-purity $N_2$ gas. Sintering temperatures of 1000-1100 °C were studied for 1-60 minutes at a heating rate of 50 °C/min. For sintering times of below 10 minutes, the samples were pushed into the furnace and pulled out of it at the maximum temperature. It took 30-35 s at temperatures above 1065 °C for the NiCrSiB powders to reach the furnace temperature and to melt into a drop shape. During this step, some of the drops spread rapidly as a classic free-flowing alloy and infiltrated into the compacts. Changes in contact angle with time were measured directly from video recordings (Fig. 1b). As shown in Fig. 1b, the BNi2 alloy wets the surface of the K05 green compact perfectly, unlike the other two alloys. The contact angles of the two hardfacing alloys soon find a steady state, but at much higher values (about 22° and 51°), mostly as a result of their different liquid fluidities when compared with BNi2. In general, eutectic points with high negative slopes of the liquidus line usually lead to highly fluent molten alloys that are appropriate for brazing. In the present study, the liquid state fluidity resulted primarily from the chemical composition of the alloy. The nearer the alloy to the eutectic region, the more fluent was the liquid. It should be noted that hardfacing alloys are usually produced with less sensitivity to their precise composition and to obtaining anamorphous state, both of which affect the molten fluidity.

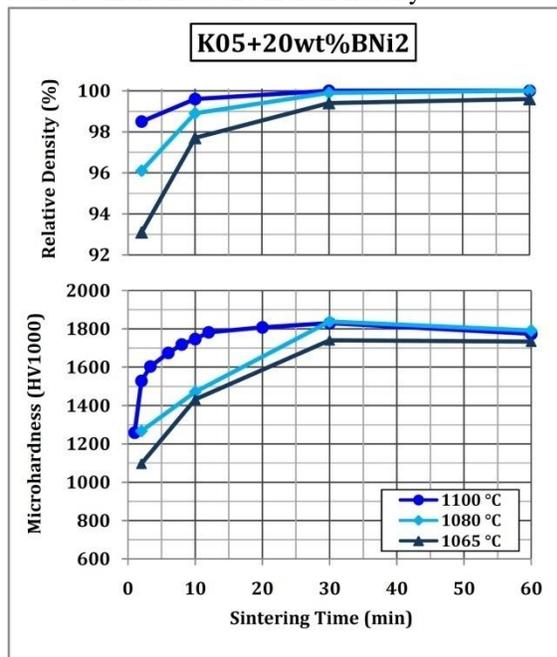

Fig. 2. Relative Density (calculated using the Archimedes method) and microhardness (HV1) of K05 samples that had been sintered *via* the infiltration of 20wt%BNi2 binder alloy at different temperatures, plotted as a function of sintering time.

The presence of carbide former elements such as Cr and Si may simplify the wetting process and may explain the difference between the wetting behaviors of the two hardfacing alloys (Ni35A and Ni22).

In Fig. 1c, consecutive frames of a sessile drop test using a short rod of BNi2 (again on a green K05 disk) are shown. The results are in accordance with those presented in Fig. 1b. BNi2 can be seen to wet the surface of the K05 green disk perfectly and to infiltrate into it completely.

Since the contact angles and relative densities of the samples that were infiltrated by the BNi2 alloy show superior behavior to the two other NiCrSiB alloys, further data and discussions in this article are focused only on these samples. In Fig. 2, the relative density (calculated using the Archimedes method) and microhardness (HV1) of the K05 samples after pressureless sintering by infiltration of the 20wt% BNi2 binder alloy at different temperatures are plotted as a function of time. It can be seen that pressureless infiltration of the green K05 samples by the BNi2 liquid at 1100 °C results in sintering of the samples in a few minutes or less. Pressureless sintering of the K05 samples during infiltration of the highly fluent BNi2 brazing alloy for only 2 and 10 minutes (in addition to 30-35 s for heating to the final temperature) leads to relative densities of 98.5% and 99.6% and isreferred to as "immediate sintering". With regard to economical and technical considerations, this demonstration of low temperature short time sintering, under no applied pressure, is a valuable practical development in itself, even in the absence of improvements in the final properties. The very high heating rate that was used here affects the melting and wetting behavior of the BNi2 alloy significantly. As mentioned by Schwartz [22], the heating rate affects the melting behavior and temperature in at least two ways, especially at high temperature. Firstly, an insufficient heating rate in the temperature range 540-925 °C can lead to oxidation of the brazing alloy, resulting in the formation of a solid shell around the molten alloy, thereby affecting its chemical composition and reducing its molten fluency. In addition, such commercial brazing alloys are usually amorphously solidified to have a precise eutectic chemical composition in the shortest possible distance [22]. Therefore, remaining at a high temperature for too long can lead to a eutectic transformation in the unstable amorphous brazing alloy and to the formation of undesirable brittle phases, such as Ni borides and silicides, in the product, with higher melting points and different chemical compositions; the former decreases the molten fluidity at the working temperature, while the latter reduces the chemical homogeneity.

Sintering the samples for 30 minutes at 1100 °C led to a relative density of truly 100%. This achievement is of great industrial importance, considering both the duration and the temperature of the process. However, the hardness is also higher than classical values for a WC-25wt%binder of submicron initial particle size (Fig. 2). At 1080 °C, full density was obtained only after 60 min of sintering. It was not achieved at all at 1065 °C.

The hardness profiles show similar trends of hardening as a function of sintering time, in accordance with the relative density plots. It is again visible that a sintering temperature of 1100 °C has a remarkable effect on hardness even after 1, 2 or several minutes. The mean microhardness values of the samples are above1525, 1600 and 1800HV1 for samples sintered for 2, 10 and 30 minutes in succession at 1100 °C. (Each microhardness value is an average of at least five indentation tests with an accuracy of ±30). It should be noted that a microhardness test using 1000gf, which leads to indentation sizes of more than 30 μm, provides average data from hundreds of WC grains in the submicron microstructure. A slight decrease in hardness is seen in samples sintered for 60 minutes (Fig. 2), perhaps as a result of the effect of WC grain growth as a softening mechanism.

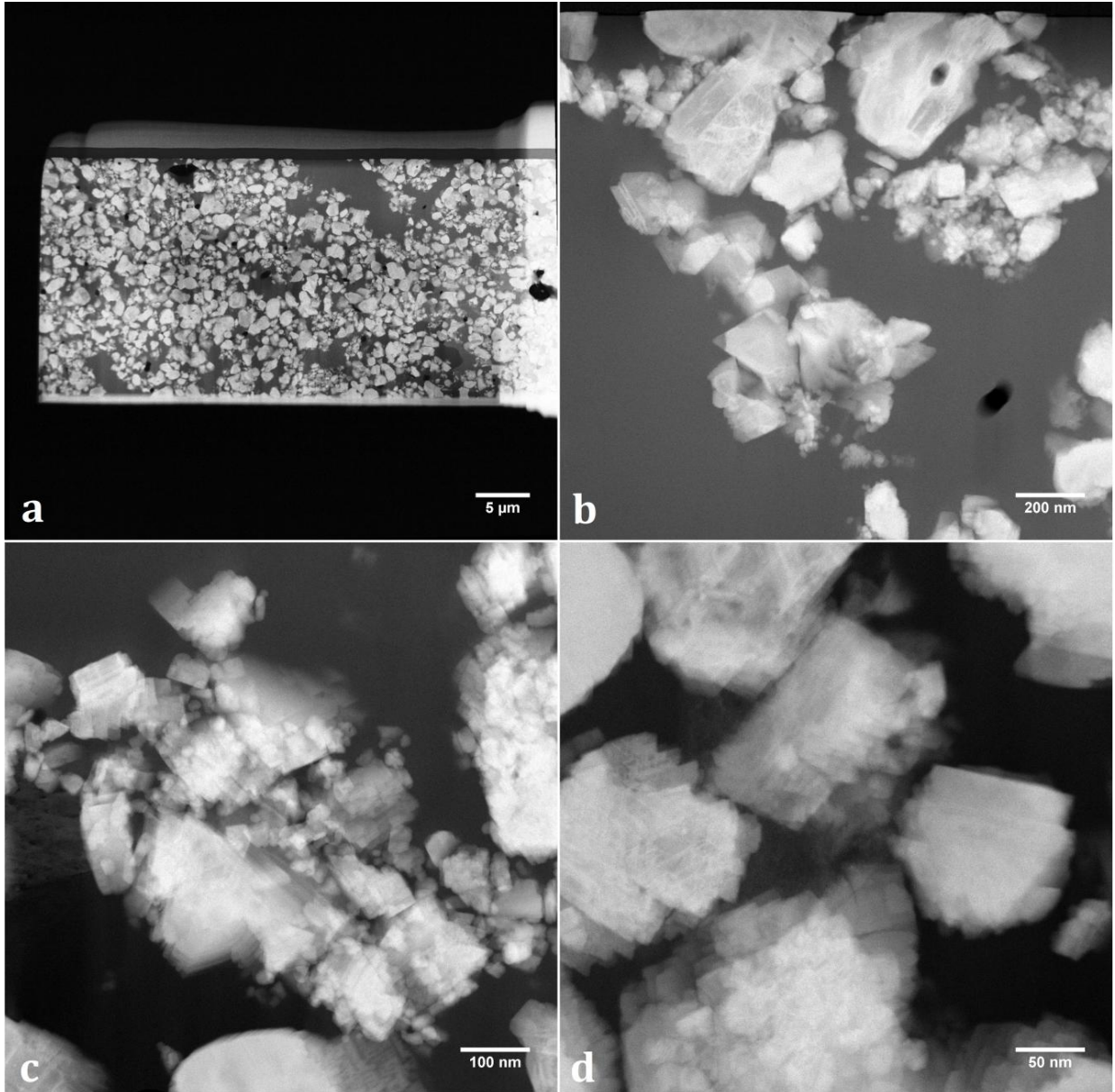

Fig. 3. HAADF images, recorded at different magnifications, of a FIB section of aK05 sample infiltrated by 20wt% BNi2 at 1100 °C for 200 s, showing perfect interface wetting (a-d), nanoparticle formation (b-d) and a stair morphology of the parent WC particles surfaces (b-d).

The microstructures of the samples were assessed using scanning transmission electron microscopy (STEM). A high-angle annular dark-field (HAADF) image of a sample that was prepared using focused ion beam (FIB) milling after sintering at 1100 $^{o}$C for 200 s is shown in Fig. 3a. About 0.5% porosity (measured using MIP® software [24]) and some heterogeneity is visible in the microstructure. Higher magnification images are shown in Fig. 3b-d. All of the interfaces show perfect wetting of the WC particles by the liquid BNi2 binder. This is a key requirement for achieving full densification and good mechanical properties. Microstructural refinement from a mean particle size of about 600 nm in the as-received precursor is visible after 200 s of sintering. A large number of nanograins, which have high crystallinity and dimensions of a few tens of nm, can be seen to have formed during the sintering process. The 10-20 nm grains are thought to have formed *in situ* from the larger WC precursors by a solution/reprecipitation mechanism, leading to a refinement of the particles from about 600 to 350-400 nm (Fig. 3b-d). This observation is in contrast to classical literature on conventional grain growth during sintering by particle coalescence [25,26].

A further observation, which is visible in Fig. 3c-d, is a sharp zigzag-shaped morphology on preferred crystallographic planes of the refined WC particles, showing either 60° (Fig. 3c) (apparently for prismatic/prismatic stairs) or 90° (Fig. 3d) (for basal/prismatic stairs). Similar observations have been reported before in the WC/Co/VC system [27–29]. This feature results from the dissolution of atoms from the surfaces of the parent WC particles [27], supporting the proposed solution/reprecipitation mechanism for nanograin formation.

As thousands of the WC particles are "brazed to each other" using a classic brazing alloy and process, we suggest to coin the word "sinterazing" by mixing the words "sintering" and "brazing" to describe this process.

In summary:
- An innovative nanostructured WC/Co/BNi2 cemented carbide was sintered by the infiltration of a BNi2 brazing alloy into the compact.
- Nanograins formed during the short sintering process.
- The cemented carbide achieves full relative density (with a microhardness of above 1800HV1) after sintering at 1100 $^{o}$C for at most 30 minutes, representing a dramatic decrease in process temperature and cost.
- Sintering at 1100 $^{o}$C for only 2 minutes leads to a relative density of 98.5% and a microhardness of more than 1500HV1.
- The final microstructure contains nanograins, which are thought to form during the sintering process by a solution/reprecipitation mechanism.

-----------------------------------------

H. Amel-Farzad gratefully acknowledges Professor A. Sabour for fruitful discussions and help both during and prior to this research.